\documentclass{revtex4}
\usepackage{graphicx}
\begin{document}
\bibliographystyle{revtex}
\preprint{FSU-HEP-2001-1019}
\title{Associated Higgs boson production with heavy quarks at Hadron
  Colliders: impact of NLO results}
\author{L. Reina}
\email[]{reina@hep.fsu.edu}
\affiliation{Physics Department, Florida State University,
Tallahassee, FL 32306-4350, USA}
\author{S. Dawson}
\email[]{dawson@quark.phy.bnl.gov}
\affiliation{Physics Department, Brookhaven National Laboratory,
Upton, NY 11973, USA}
\author{D. Wackeroth}
\email[]{dow@pas.rochester.edu}
\affiliation{Department of Physics and Astronomy, University of Rochester,
Rochester, NY 14627-0171, USA}
\date{\today}
\begin{abstract}
  We emphasize the role that the associated production of a Higgs
  boson with a pair of top-antitop quarks can play at present and
  future hadron colliders.  Results of recent calculations of the NLO
  total cross section for the associated production of a Standard
  Model like Higgs boson with a pair of top-antitop quarks at the
  Tevatron ($\sqrt{s}\!=\!2$~TeV) are presented.
\end{abstract}
%\pacs{14.80.Bn,12.38.Bx,12.15.-y,13.85.-t}
\maketitle
\section{Introduction}
\label{sec:intro}
The possibility of discovering a Higgs boson in the range between
$115-130$~GeV is becoming increasingly likely.  The Standard Model
(SM) precision fits are consistent with a light Higgs boson
\cite{Lepewwg:2001}. At the same time, the Higgs sector of the Minimal
Supersymmetric Standard Model (MSSM) requires the existence of a
scalar Higgs boson lighter than about 130~GeV
\cite{Heinemeyer:1998np}. Both the Fermilab Tevatron and the CERN
Large Hadron Collider (LHC) will focus on the search for a light Higgs
boson. Since in the low mass range, below the $W$-pair threshold, a
Higgs boson mainly decays hadronically into $b\bar b$ pairs, both the
Tevatron and the LHC will have to optimize their search strategies in
order to overcome the overwhelming hadronic background. This implies
that all available Higgs boson production and decay channels have to
be considered.

In this context, the associated production of a Higgs boson with a
pair of top-antitop quarks has drawn increasing attention. In spite of
the very small cross section, this production mode has an extremely
distinctive signature, and recent analyses have shown that it can be
within the reach not only of the LHC, but also of the Tevatron, if
integrated luminosities of 15-30~fb$^{-1}$ become available
\cite{Goldstein:2000bp}. From ongoing studies \cite{incandela:2001} we
learn that including $p\bar p\rightarrow t\bar t H$ among the Higgs
search channels could lower the luminosity required for discovery of a
SM like Higgs at the Tevatron by as much as 15-20\%, given the high
significance of the corresponding signature \cite{hobbs:2001}.  If not
at the Tevatron, this mode will surely be used at the LHC, where it is
instrumental in the discovery of a light SM like Higgs
\cite{TDRCMS:1994,TDRATLAS:1999}. The higher statistics available at
the LHC will also allow a first precision measurement of the top quark
Yukawa coupling at the 20\% level.

In view of the relevance that this production mode can have in
particular for the Tevatron, we have recently completed the
calculation of the inclusive total cross section for $p\bar
p\rightarrow t\bar t h$, for a SM Higgs ($h\!=\!h_{SM}$), at the
Tevatron center of mass energy $\sqrt{s}\!=\!2$~TeV, including first
order QCD corrections \cite{Reina:2001sf,Reina:2001bc}. The main
impact of next-to-leading order (NLO) QCD corrections is in reducing
the dependence on the renormalization and factorization scales of the
Born level cross section enormously, giving us increased confidence in
our theoretical predictions. We will briefly discuss the
characteristics of our calculation in Sec.~\ref{sec:general}, and
present our results in Sec.~\ref{sec:results}. This calculation has
also been performed by the authors of Ref.~\cite{Beenakker:2001rj},
and results of our two groups are in very good agreement.

\section{General Framework}
\label{sec:general}

The ${\cal O}(\alpha_s^3)$ total cross section for $p\bar p\rightarrow
t\bar t h$ is defined as:
\begin{equation}
\label{eq:sigma_hadr}
\sigma_{NLO}(p\bar p\rightarrow t\bar t h)=
\sum_{ij}\int dx_1 dx_2 
{\cal F}_i^p(x_1,\mu) {\cal F}_j^{\bar p}(x_2,\mu)
{\hat \sigma}^{ij}_{NLO}(x_1,x_2,\mu)\,\,\,,
\end{equation}
where we denote by ${\cal F}_i^{p,\bar p}$ the NLO parton distribution
functions for parton $i$ in a proton/antiproton, defined at a generic
factorization scale $\mu_f\!=\!\mu$. ${\hat \sigma}^{ij}_{NLO}$ is the
${\cal O}(\alpha_s^3)$ parton level total cross section for incoming
partons $i$ and $j$, made of the two channels $q\bar q,\,
gg\rightarrow t\bar t h$, and renormalized at an arbitrary scale
$\mu_r$ which we also take to be $\mu_r\!=\!\mu$.  We note that
because of the large mass of the produced $t\bar th$ final state, this
process is very close to threshold at the Tevatron, for $p\bar p$
collision at center of mass energy $\sqrt{s}\!=\!2$~TeV.  As a
consequence, at the Tevatron more than $95\%$ of the tree level total
cross section comes from $q\bar q\rightarrow t\bar th$, summed over
all light quark flavors, and the $gg$ contribution is completely
negligible. Therefore we compute $\sigma_{NLO}(p\bar p\rightarrow
t\bar th)$ by including in ${\hat \sigma}^{ij}_{NLO}$ only the ${\cal
  O}(\alpha_s)$ corrections to $ q\bar q\rightarrow t\bar t h$.  The
calculation of $g g \rightarrow t\bar t h$ at ${\cal O}(\alpha_s^3)$
is, however, crucial to determine $\sigma_{NLO}(pp\rightarrow t \bar
th)$ for the LHC, since in $pp$ collisions at $\sqrt{s}\!=\!14$~TeV a
large fraction of the total cross section comes from the
$gg\rightarrow t\bar th$ channel.  The ${\cal O}(\alpha_s^3)$ total
cross section for the LHC has been estimated within the Effective
Higgs Approximation in Ref.~\cite{Dawson:1998im}.  Full results are
presented in Ref.~\cite{Beenakker:2001rj} and will also appear in
Ref.~\cite{Dawson:2001gg}.

The ${\cal O}(\alpha_s^3)$ parton level total cross section
can be written as:
\begin{equation}
\label{eq:sigma_part}
{\hat \sigma}_{NLO}^{ij}(x_1,x_2,\mu)=
\alpha_s^2(\mu)\biggl\{{\hat f}_{LO}^{ij}(x_1,x_2)+
{\alpha_s(\mu)\over 4 \pi}{\hat f }_{NLO}^{ij}(
x_1,x_2,\mu)\biggr\}
\equiv{\hat \sigma}_{LO}^{ij}(x_1,x_2,\mu)+
\delta {\hat \sigma}_{NLO}^{ij}(x_1,x_2,\mu)\,\,\,,
\end{equation}
where $\delta {\hat \sigma}_{NLO}^{ij}( x_1,x_2,\mu)$ consists of both
${\cal O}(\alpha_s)$ virtual and real corrections to the Born cross
section :
\begin{equation}
\label{eq:sigma_virt_real}
\delta {\hat \sigma}_{NLO}^{ij}(
x_1,x_2,\mu)=\hat\sigma_{virt}^{ij}+\hat\sigma_{real}^{ij}\,\,\,.
\end{equation}
The virtual part of the NLO cross section contains UV divergences
that are renormalized by introducing a suitable set of counterterms.
It also contains IR singularities that are cancelled by analogous
singularities in the real part of the NLO cross section and in the
renormalized parton distribution functions.

The calculation of the ${\cal O}(\alpha_s)$ virtual corrections has
required us to evaluate pentagon scalar integrals with several massive
external and internal particles. These integrals were not available in
the literature, and we have calculated them by reducing them to a
linear combination of box scalar integrals, by applying the method
originally introduced in \cite{Bern:1993em,Bern:1994kr}.

The ${\cal O}(\alpha_s)$ real corrections, i.e. the cross section for
$q\bar q\rightarrow t\bar th+g$, have been calculated using two
different implementations of the Phase Space Slicing (PSS) method, with
the introduction of one \cite{Giele:1992vf,Giele:1993dj,Keller:1998tf}
or two cutoffs \cite{Harris:2001sx} respectively.  In both cases, the
IR singularities due to the emission of either a soft or a collinear
gluon can be isolated in specific regions of the phase space, and
calculated analytically, while the integration over the remaining
phase space is performed numerically using standard Monte Carlo
techniques. In fact, this is the first application of the one cutoff PSS
method to a case with more than one massive particle in the final
state. A detailed description of our calculation can be found in
Ref.~\cite{Reina:2001bc}.

\section{Results}
\label{sec:results}

All the results presented in this section have been obtained using
CTEQ4M parton distribution functions ~\cite{Lai:1997mg} and the 2-loop
evolution of $\alpha_s(\mu)$ for the calculation of the NLO cross
section, and CTEQ4L parton distribution functions and the 1-loop
evolution of $\alpha_s(\mu)$ for the calculation of the lowest order
cross section, $\sigma_{LO}$. The top-quark mass is taken to be
$m_t\!=\!174$~GeV and $\alpha_s^{NLO}(M_Z)\!=\!0.116$.

\begin{figure}
  \begin{minipage}[b]{.46\linewidth}
  \includegraphics[scale=.45]{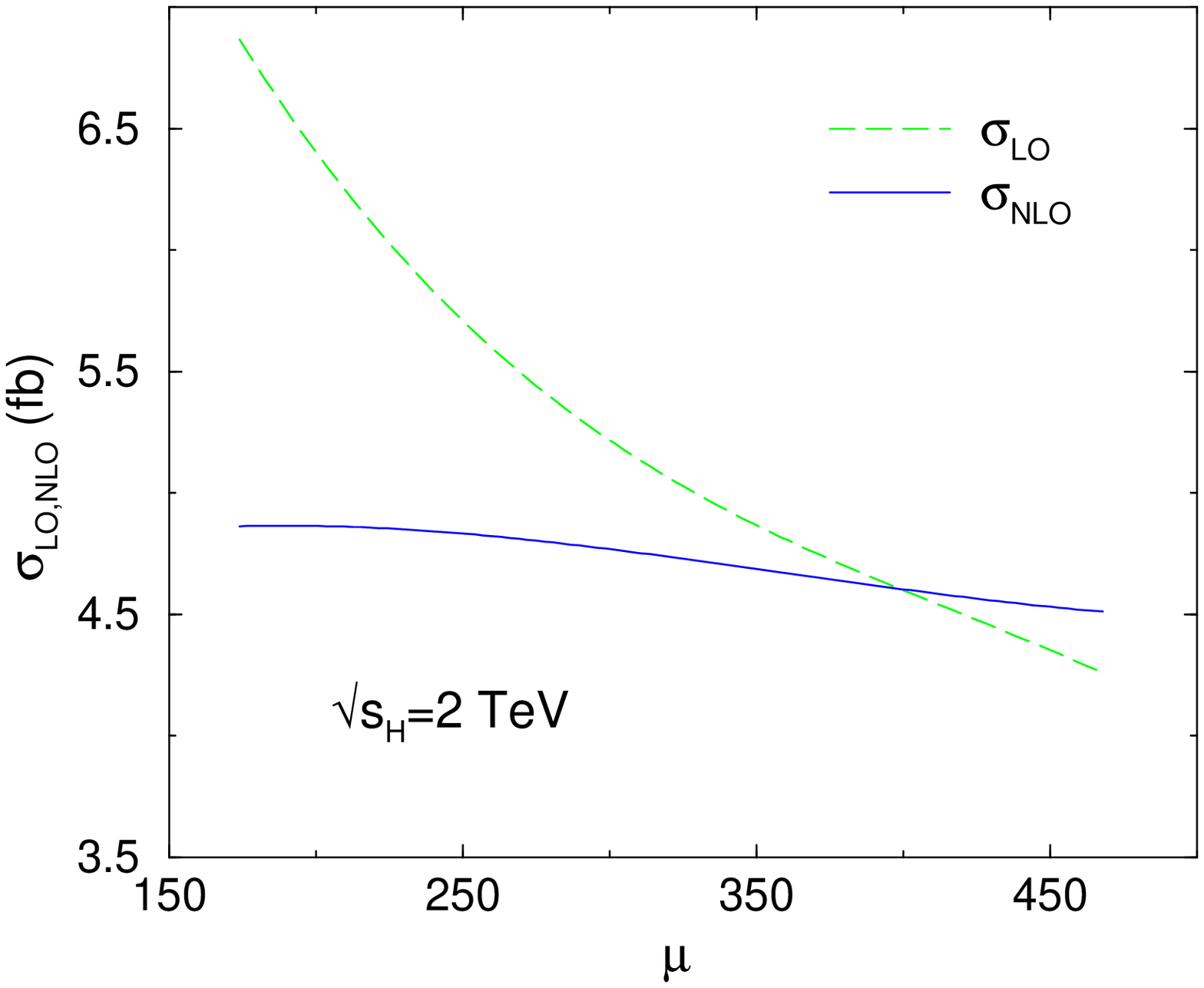}
  \caption{Dependence of $\sigma_{LO,NLO}(p\bar p\rightarrow
   t\bar t h)$ on the renormalization scale $\mu$, at
   $\sqrt{s}\!=\!2$~TeV, for $M_h\!=\!120$~GeV. }
  \label{fig:mudep}
  \end{minipage}\hspace{0.5truecm}
  \begin{minipage}[b]{.46\linewidth}
  \includegraphics[scale=.45]{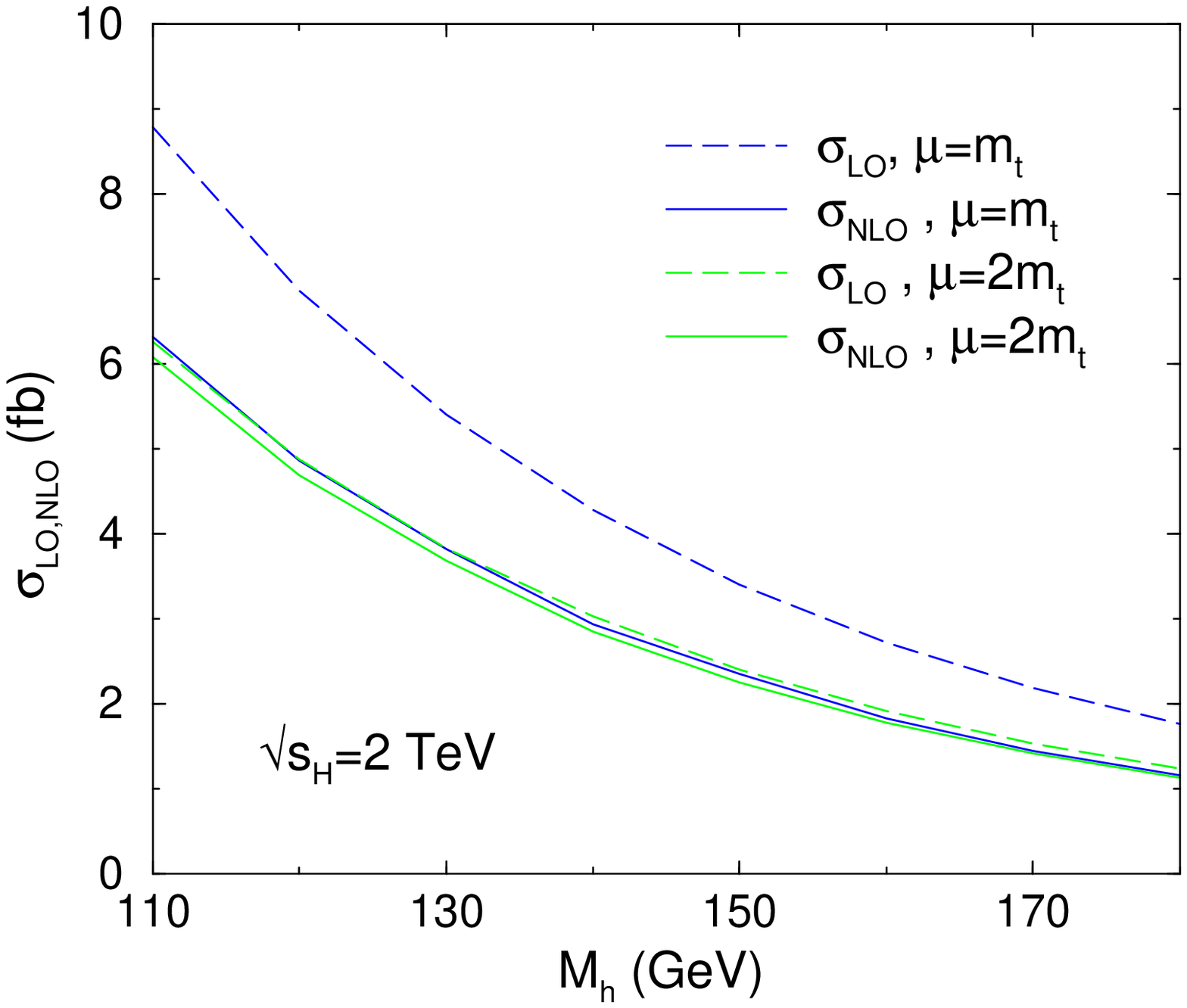}
  \caption{
  $\sigma_{NLO}$ and $\sigma_{LO}$ for $p\bar p\rightarrow t\bar t h$
  as functions of $M_h$, at $\sqrt{s}\!=\!2$~TeV, for $\mu\!=m_t$ and
  $\mu\!=\!2m_t$.}
  \label{fig:signlo}
  \vspace{0.3truecm}
  \end{minipage}
 \end{figure}

The importance of having calculated the total cross section for $p\bar
p\rightarrow t\bar t h$ at the NLO of QCD corrections is manifest in
Fig.~\ref{fig:mudep}, where we show, for $M_h\!=\!120$~GeV, how at NLO
the dependence on the arbitrary renormalization/factorization scale
$\mu$ is significantly reduced. For $M_h\!=\!120$~GeV, the NLO
cross section varies in the range $4.8-4.5$~fb with a residual
renormalization and factorization scale dependence of the order of
$8\%$. For larger Higgs masses the cross section becomes much smaller,
with values of the order of 1~fb for $M_h\!=\!180$~GeV, as illustrated
in Fig.~\ref{fig:signlo}. Combining the residual scale dependence with
the error from the parton distribution functions (6\%) and from $m_t$
(7\%), we estimate the uncertainty on our theoretical prediction at
about 12\%.
 
\begin{figure}
\scalebox{0.5}{\includegraphics{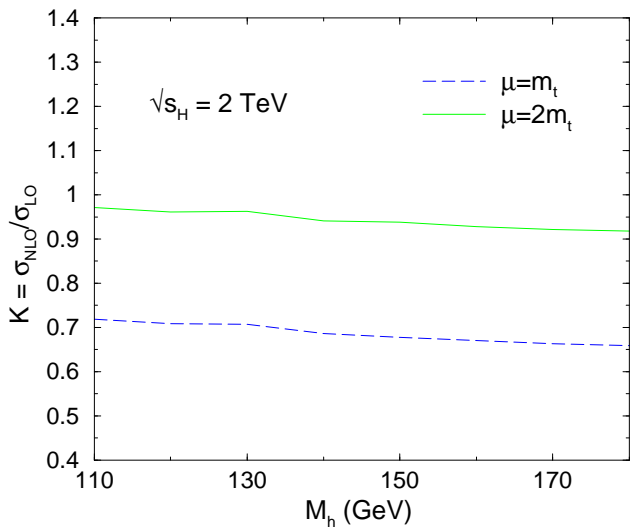}}
  \caption{K factor for $p
   {\overline p}\rightarrow t {\overline t} h$ as a function of $M_h$,
   at $\sqrt{s}\!=\!2$~TeV, for $\mu\!=m_t$ and
   $\mu\!=\!2m_t$.}
\label{fig:kfac}
\end{figure}

 In Fig.~\ref{fig:mudep}, we also notice that the LO and NLO cross
 section curves cross at a scale between $2 m_t$ and $2 m_t+M_h$. If
 we define as customary a K-factor as the ratio between the NLO and
 the LO cross sections, $K=\sigma_{NLO}/\sigma_{LO}$, the K factor for
 $p\bar p\rightarrow t\bar th$ turns out to be smaller than one for
 scales roughly below $2m_t+M_h$ and bigger than one otherwise. The
 dependence of the K-factor on $M_h$ is very mild, as shown in
 Fig.~\ref{fig:kfac}, where we plot the behavior of the K-factor for
 scales $\mu\!=\!m_t$ ($K\!\simeq\! 0.7$) and $\mu\!=\!2m_t$
 ($K\!\simeq\! 0.95$). It is worth noting, however, that, given the
 strong scale dependence of the LO cross section, the K-factor also
 shows a significant $\mu$-dependence and therefore is an equally
 unreliable prediction.  Therefore we would like to stress once more
 that we only discuss the K-factor as a qualitative indication of the
 impact of ${\cal O}(\alpha_s)$ QCD corrections. The physically
 meaningful quantity is the NLO cross section, not the K-factor.

\section{Conclusions}
\label{sec:concl}
The NLO inclusive total cross section for the Standard Model process
$p\bar p\rightarrow t\bar th$ at $\sqrt{s}\!=\!2$~TeV shows a
drastically reduced scale dependence as compared to the Born result
and leads to increased confidence in predictions based on these
results.  The NLO QCD corrections slightly decrease or increase the
Born level cross section depending on the renormalization and
factorization scales used. The NLO inclusive total cross section for
Higgs boson masses in the range accessible at the Tevatron,
$120\!<\!M_h\!<\!180$ GeV, is of the order of $1-5$ fb.

\begin{acknowledgments}
  The work of L.R. (S.D.)  is supported in part by the U.S. Department
  of Energy under grant DE-FG02-97ER41022 (DE-AC02-76CH00016). The
  work of D.W.  is supported by the U.S. Department of Energy under
  grant DE-FG02-91ER40685.
\end{acknowledgments}

\bibliography{tth_snow}

\begin{thebibliography}{10}
\providecommand*{\bibinfo}[2]{#2}
\providecommand*{\eprint}[1]{#1}
\providecommand*{\url}[1]{#1}
\bibitem{Lepewwg:2001}
\bibinfo{author}{LEPEWWG/2001-01} (\bibinfo{date}{May 2001}).
\bibitem{Heinemeyer:1998np}
\bibinfo{author}{S.~Heinemeyer}, \bibinfo{author}{W.~Hollik}, and
  \bibinfo{author}{G.~Weiglein}, \bibinfo{journal}{Eur. Phys. J.}
  \bibinfo{volume}{\textbf{C9}}, \bibinfo{pages}{343} (\bibinfo{date}{1999}),
  \eprint{arXiv:hep-ph/9812472}.
\bibitem{Goldstein:2000bp}
\bibinfo{author}{J.~Goldstein} \emph{et~al.}, \bibinfo{journal}{Phys. Rev.
  Lett.} \bibinfo{volume}{\textbf{86}}, \bibinfo{pages}{1694}
  (\bibinfo{date}{2001}), \eprint{arXiv:hep-ph/0006311}.
\bibitem{incandela:2001}
\bibinfo{author}{J.~Incandela}, Fermilab (\bibinfo{date}{May 3-5 2001}), talk
  presented at the \emph{Workshop on the Future of Higgs Physics}.
\bibitem{hobbs:2001}
\bibinfo{author}{J.~Hobbs} (\bibinfo{date}{July 2001}), talk presented at
  \emph{Snowmass 2001}.
\bibitem{TDRCMS:1994}
\bibinfo{title}{\emph{CMS Collaboration}}, Tech. Rep. CERN/LHCC/94-38, CERN
  (\bibinfo{date}{1994}).
\bibitem{TDRATLAS:1999}
\bibinfo{title}{\emph{ATLAS Collaboration}}, Tech. Rep. CERN/LHCC/99-15, CERN
  (\bibinfo{date}{1999}).
\bibitem{Reina:2001sf}
\bibinfo{author}{L.~Reina} and \bibinfo{author}{S.~Dawson}
  (\bibinfo{date}{2001}), \eprint{arXiv:hep-ph/0107101}.
\bibitem{Reina:2001bc}
\bibinfo{author}{L.~Reina}, \bibinfo{author}{S.~Dawson}, and
  \bibinfo{author}{D.~Wackeroth} (\bibinfo{date}{2001}),
  \eprint{arXiv:hep-ph/0109066}.
\bibitem{Beenakker:2001rj}
\bibinfo{author}{W.~Beenakker} \emph{et~al.} (\bibinfo{date}{2001}),
  \eprint{arXiv:hep-ph/0107081}.
\bibitem{Dawson:1998im}
\bibinfo{author}{S.~Dawson} and \bibinfo{author}{L.~Reina},
  \bibinfo{journal}{Phys. Rev.} \bibinfo{volume}{\textbf{D57}},
  \bibinfo{pages}{5851} (\bibinfo{date}{1998}), \eprint{arXiv:hep-ph/9712400}.
\bibitem{Dawson:2001gg}
\bibinfo{author}{S.~Dawson}, \bibinfo{author}{L.~Orr},
  \bibinfo{author}{L.~Reina}, and \bibinfo{author}{D.~Wackeroth}, work in
  progress.
\bibitem{Bern:1993em}
\bibinfo{author}{Z.~Bern}, \bibinfo{author}{L.~J. Dixon}, and
  \bibinfo{author}{D.~A. Kosower}, \bibinfo{journal}{Phys. Lett.}
  \bibinfo{volume}{\textbf{B302}}, \bibinfo{pages}{299} (\bibinfo{date}{1993}),
  \eprint{arXiv:hep-ph/9212308}.
\bibitem{Bern:1994kr}
\bibinfo{author}{Z.~Bern}, \bibinfo{author}{L.~J. Dixon}, and
  \bibinfo{author}{D.~A. Kosower}, \bibinfo{journal}{Nucl. Phys.}
  \bibinfo{volume}{\textbf{B412}}, \bibinfo{pages}{751} (\bibinfo{date}{1994}),
  \eprint{arXiv:hep-ph/9306240}.
\bibitem{Giele:1992vf}
\bibinfo{author}{W.~T. Giele} and \bibinfo{author}{E.~W.~N. Glover},
  \bibinfo{journal}{Phys. Rev.} \bibinfo{volume}{\textbf{D46}},
  \bibinfo{pages}{1980} (\bibinfo{date}{1992}).
\bibitem{Giele:1993dj}
\bibinfo{author}{W.~T. Giele}, \bibinfo{author}{E.~W.~N. Glover}, and
  \bibinfo{author}{D.~A. Kosower}, \bibinfo{journal}{Nucl. Phys.}
  \bibinfo{volume}{\textbf{B403}}, \bibinfo{pages}{633} (\bibinfo{date}{1993}),
  \eprint{arXiv:hep-ph/9302225}.
\bibitem{Keller:1998tf}
\bibinfo{author}{S.~Keller} and \bibinfo{author}{E.~Laenen},
  \bibinfo{journal}{Phys. Rev.} \bibinfo{volume}{\textbf{D59}},
  \bibinfo{pages}{114004} (\bibinfo{date}{1999}),
  \eprint{arXiv:hep-ph/9812415}.
\bibitem{Harris:2001sx}
\bibinfo{author}{B.~W. Harris} and \bibinfo{author}{J.~F. Owens}
  (\bibinfo{date}{2001}), and references therein,
  \eprint{arXiv:hep-ph/0102128}.
\bibitem{Lai:1997mg}
\bibinfo{author}{H.~L. Lai} \emph{et~al.}, \bibinfo{journal}{Phys. Rev.}
  \bibinfo{volume}{\textbf{D55}}, \bibinfo{pages}{1280} (\bibinfo{date}{1997}),
  \eprint{arXiv:hep-ph/9606399}.

\end{thebibliography}
\end{document}